# Capacitance of Silicon Pixels

Grant Gorfine, Martin Hoeferkamp, Geno Santistevan, and Sally Seidel

New Mexico Center for Particle Physics
University of New Mexico
Albuquerque, New Mexico 87131 U.S.A.

**Abstract**

Capacitance measurements have been made on silicon pixel sensors of types $n^+$ on $n$, $p^+$ on $n$, and $n^+$ on $p$. The arrays test a variety of implant and gap widths, and the $n^+$ on $n$ devices test several p-stop designs. The measurements examine inter-pixel and backplane contributions and include studies of temperature dependence. Measurements were made before and after irradiation with fluences relevant to LHC experiments and Fermilab Tevatron Run 2.

# 1 Introduction

The capacitance is a sensitive parameter in the operation of a silicon tracking detector, as it directly affects the noise and crosstalk. While the capacitance of silicon strip detectors has been previously studied[1], the small feature size of silicon pixel detectors has rendered measurements of their capacitance significantly more challenging. It is expected that the capacitance of each pixel cell includes contributions from all of its neighbors in the array. Both the inter-pixel and backplane capacitance are expected to contribute substantially to the total capacitance of the sensor; consequently, inter-pixel and pixel backplane capacitance measurements for a variety of designs are necessary. The pixel detectors in ATLAS[2] will be operated cold ($-6°$C), as will those in other upcoming experiments, so studies have been made at various temperatures to understand the temperature dependence of the pixel capacitance.

Pixel detectors at the LHC or the Tevatron will be subject to high radiation levels[2]. The second barrel layer of ATLAS, located at a radius of about 10 cm from the interaction point, is conservatively expected to receive a fluence equivalent to $10^{15}$ 1 MeV neutrons cm$^{-2}$ during 10 years of operation. Inter-strip capacitance is known to increase with irradiation prior to inversion [3, 4]. It is consequently important to investigate the effects of irradiation on the inter-pixel capacitance.

The total capacitance load presented by a pixel to the front-end electronics includes several components. One of them is the inter-pixel capacitance, which is dominated by the contributions between the pixel and its eight nearest neighbors in the array. A second is the capacitance between the pixel and the backplane. (We call this the backplane capacitance.) In addition to these, a pixel detector bump bonded to a front-end readout electronics chip receives a capacitive load from the bump bond ($\approx 40-50$ fF) and the preamplifier input transistor ($\approx 60-100$ fF). We present here measurements of the inter-pixel and backplane capacitances.

The study presented here was organized as follows. We began by studying a set of pixel test structures (the "LBNL Test Structures") whose design is simple enough that measurement of the capacitance between a pixel and all of its nearest neighbors can be made unambiguously. The design of these arrays allowed us to represent them straightforwardly in electrostatic simulations. Simulations (both two- and three-dimensional) were carried out and shown to agree well with capacitance measurements on these arrays. The agreement between simulation and measurement confirms that calibrations and systematic effects in the measurements are correctly understood. There are two varieties of LBNL test structures, one with $n^+$ implants in $p$-type substrate and the other with $p^+$ implants in $n$-type substrate.

We next studied a set of $n^+$ on $n$-type arrays ("Structure 6") whose design more closely resembles that of a typical pixel detector in a particle physics experiment. As the measurement and calibration procedure and the systematic effects associated with these studies are identical to those in the LBNL test structure studies, simulations were not necessary to validate those results.

The ultimate goal of these studies is the measurement of the capacitance of several devices including the ATLAS production pixel sensors and prototypes proposed for use at Fermilab Tevatron Run 2. The former will be presented in a companion paper to this one[5].

## 2  Description of the Test Structures

### 2.1  LBNL Test Structures

Figure 1 shows a cross section of one of the LBNL $p$-type test structures, and Figure 2 shows the details of one of its pixel arrays. The test structures include six $3 \times 9$ arrays of pixels. The center pixel in each array is isolated, and the surrounding pixels are connected together. This allows all the surrounding pixels to be biased with a single probe. The pitch is 50 $\mu$m $\times$ 536 $\mu$m. The $p$-bulk detectors have a $p^+$ isolation implant (also know as a p-stop)[6] surrounding the $n^+$ implants. The widths of the pixel implants ($p^+$ in the case of $n$-type bulk and $n^+$ in the case of $p$-type bulk) are systematically varied for each array. For the $p$-type detectors, the p-stop widths and the gaps (unimplanted regions) between the $n^+$ implants and p-stops are also varied in the six arrays.

The values of the pixel implant width (denoted as $W$), the p-stop width (denoted as $P$), and the gap between p-stop and $n^+$ implant are listed in Table 1 for each of the arrays. There is also an unimplanted region 100 $\mu$m wide between the edge of the array and the guard ring, which is 30 $\mu$m wide. The thickness of the detectors is 300 $\mu$m. A total of five $p$-type and five $n$-type LBNL test structures were studied. We have chosen to show representative results in the figures of this paper rather than the complete dataset.

**Table 1** Dimensions of the LBNL test structure pixel arrays, in microns.

| Array Number | $n$-type | | $p$-type | | |
|---|---|---|---|---|---|
| | $p^+$ Implant Width | Gap Width | $n^+$ Implant Width | p-stop Width | Gap Width |
| 2 | 38 | 12 | 38 | 4 | 4 |
| 3 | 32 | 18 | 32 | 6 | 6 |
| 4 | 23 | 27 | 23 | 8 | 8 & 11[‡] |
| 5 | 20 | 30 | 20 | 10 | 10 |
| 6 | 14 | 36 | 14 | 12 | 12 |

[‡]In $p$-type Array 4, there was an error in the layout that caused the gap to be 8 $\mu$m wide on one long side and 11 $\mu$m wide on the other.

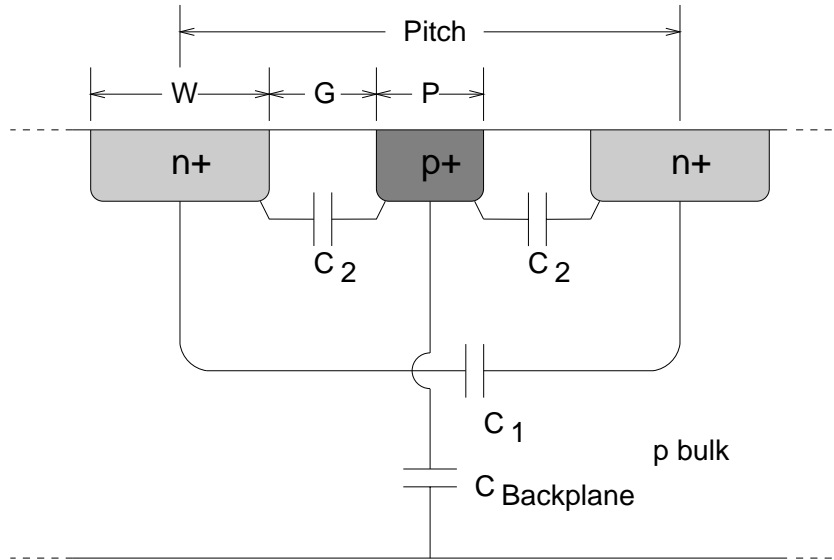

Figure 1: Schematic cross section of an LBNL $p$-type test structure illustrating the definitions of W, P, and G as well as contributions to the capacitance.

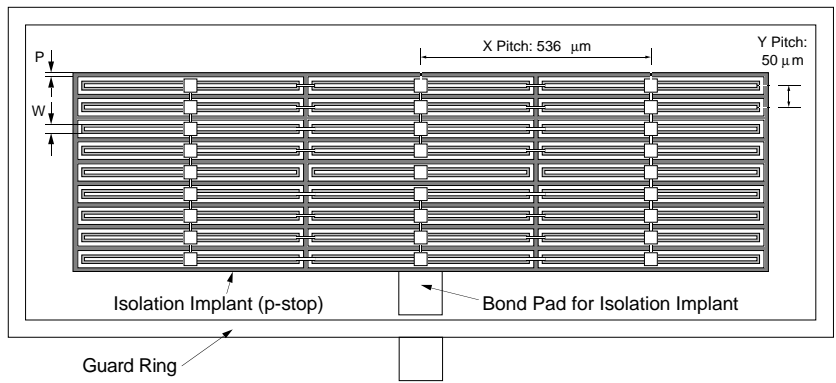

Figure 2: One pixel array in the LBNL $p$-type test structure.

## 2.2 Structure 6

A test structure designated "Structure 6" was included in the first ATLAS pixel sensor prototype submission. This test structure, nominally 300 $\mu$m thick, contains 11 arrays, each with variations in implant widths and p-stop design. Table 2 summarizes these features. Each array includes 33 pixels in a 3 × 11 matrix. The center three pixels are isolated while the surrounding ones are all connected together. A typical Structure 6 array is shown in Figure 3.

**Table 2** Summary of the dimensions of the Structure 6 arrays, in microns.

| Array No. | p-stop Design | W | A | P | G | H |
|---|---|---|---|---|---|---|
| 1 | Atoll | 23 | 18 | 5 | 6 | 5 |
| 2 | Atoll | 23 | 29 | 5 | 6 | 5 |
| 3 | Atoll | 16 | 12 | 5 | 6 | 12 |
| 4 | Atoll | 15 | 12 | 5 | 10 | 5 |
| 5 | Atoll | 19 | 14 | 5 | 8 | 5 |
| 6 | Combined | 13 | 12 | 5 | 6 | 5 |
| 7 | Common | 33 | 25 | 5 | 6 | N/A |
| 8 | Common | 28 | 23 | 10 | 6 | N/A |
| 9 | Common | 23 | 18 | 15 | 6 | N/A |
| 10 | Common | 24 | 19 | 10 | 8 | N/A |
| 11 | Common | 20 | 15 | 10 | 10 | N/A |

Definitions of the symbols used in the table:
W = $n^+$ implant width
A = metal width
P = p-stop width
G = Gap between $n^+$ and $p^+$ implants
H = Gap between neighboring $p^+$ implants

Three p-stop designs are included. One is the "atoll" design which has a p-stop ring around each pixel. The second is the "common" design, in which a continuous p-stop forms a grid around the pixels. The third is the "combined" p-stop design, which uses both atoll and common p-stops. Figure 4 shows the detail of the corner of one of the arrays utilizing the atoll p-stop. Figure 5 shows the common p-stop, and Figure 6, the combined. All five of the Structure 6 devices studied have single metal readout and are $n^+$-on-$n$.

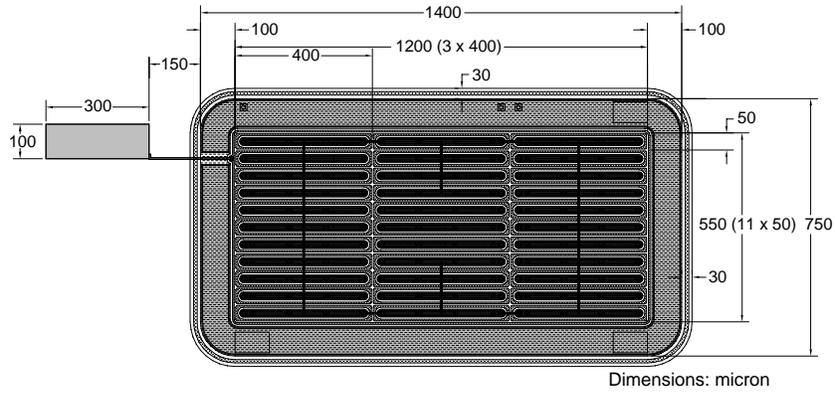

Figure 3: One pixel array on Structure 6. All dimensions are in microns.

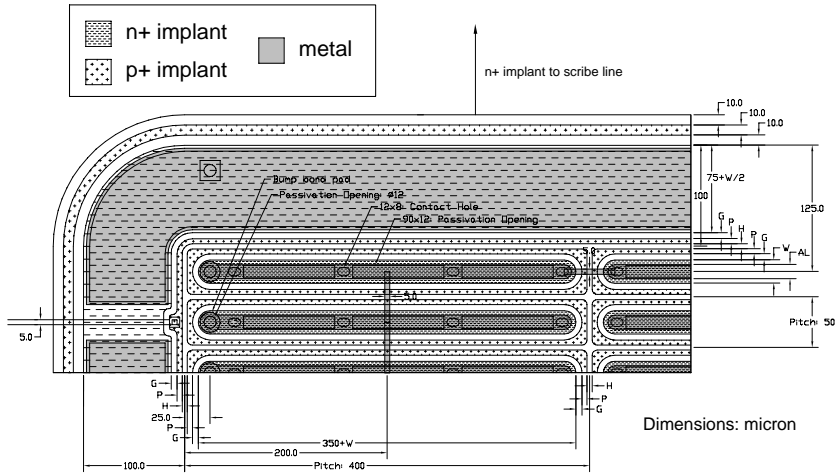

Figure 4: Detail of a corner of a pixel array on Structure 6 showing the atoll design p-stop. All dimensions are in microns.

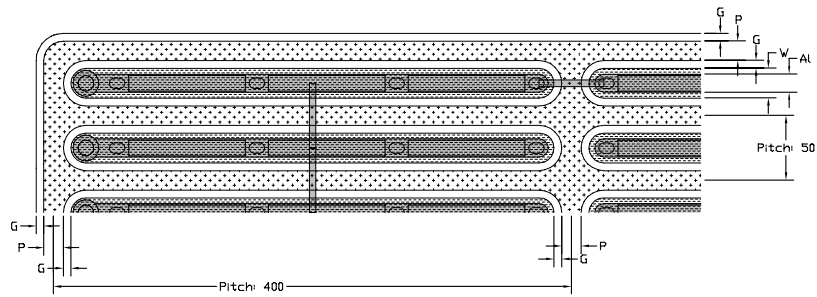

Figure 5: Detail of a corner of a pixel array on Structure 6 showing the common design p-stop. All dimensions are in microns. The symbol key is shown in Figure 4.

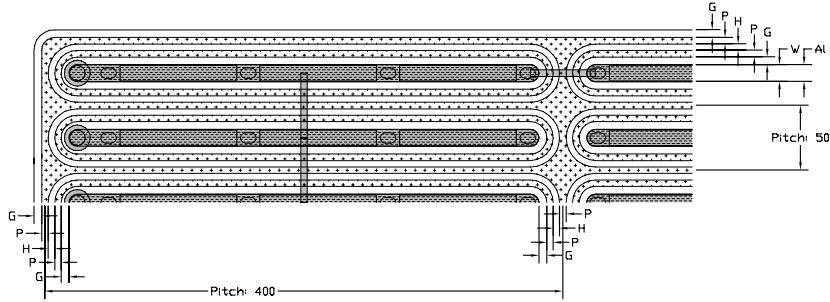

Figure 6: Detail of a corner of a pixel array on Structure 6 showing the combined design p-stop. All dimensions are in microns. The symbol key is shown in Figure 4.

## 3 Measurement Setup

The very low capacitances involved mandate that a variety of precautions be taken when measuring the inter-pixel capacitance. The test setup must be adequately shielded, and low parasitic capacitance probes are essential. Capacitance measurements were made with an HP4284A LCR meter. This supplies a small AC test signal (set to 250 mV RMS) on the HIGH terminal. The amplitude and phase are measured on the LOW terminal. Figure 7 shows a schematic of the measurement setup, and Figure 8 shows a block diagram of the probe station setup. Figure 8 shows the optional use of a temperature chamber which was required for measurements reported in Section 4.2.5 and certain studies of irradiated sensors. To measure the inter-pixel capacitance between a pixel of interest and all or some of the surrounding pixels, the pixel of interest is connected to the LOW terminal via a low parasitic capacitance probe, and the HIGH terminal is connected to the neighbors of interest (either by several probes or by having the surrounding pixels connected together). The LBNL and Structure 6 devices have several pixels connected together, a feature which facilitated this measurement.

Prior to taking a measurement, an OPEN correction is done on the HP4284A LCR meter. This procedure consists in raising the probe attached to the LOW terminal a few microns above the pixel under test. The sensor is biased, typically to 100 V. The meter is then placed in OPEN correction mode and measures all the parasitic capacitances in the setup, including cable and probe contributions. The measurement is stored and used as the reference point in subsequent measurements.

The statistical error on any particular capacitance measurement reported here is 3 fF, a value taken from the standard deviation of a distribution of identical measurements made on a typical unirradiated test structure at a representative frequency (1 MHz) and voltage (200 V). As a systematic error we take 1 fF, which is a conservative measure of the voltage dependence of the OPEN correction on a typical capacitance measurement. The residual parasitic capacitance after completion of the OPEN correction was less than 2 fF. The accuracy of the capacitance measurement, as given by the LCR meter specifications, is 3 fF. We combine these three errors in quadrature to get 5 fF as the total measurement error.

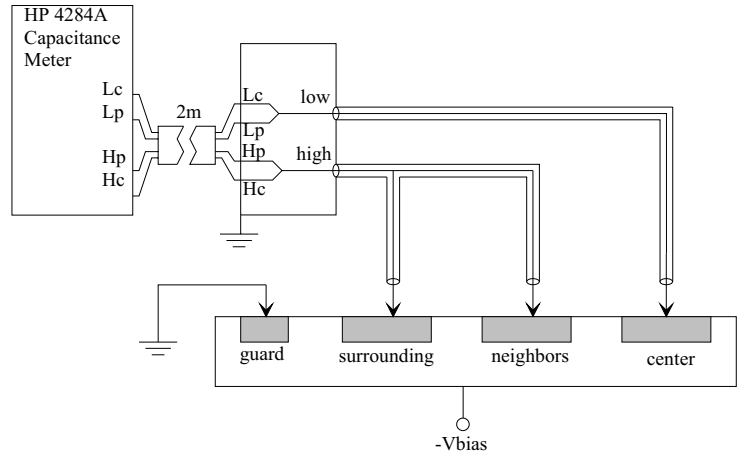

Figure 7: Schematic of the setup for the inter-pixel capacitance measurement. The capacitance is measured between the HIGH and LOW terminals.

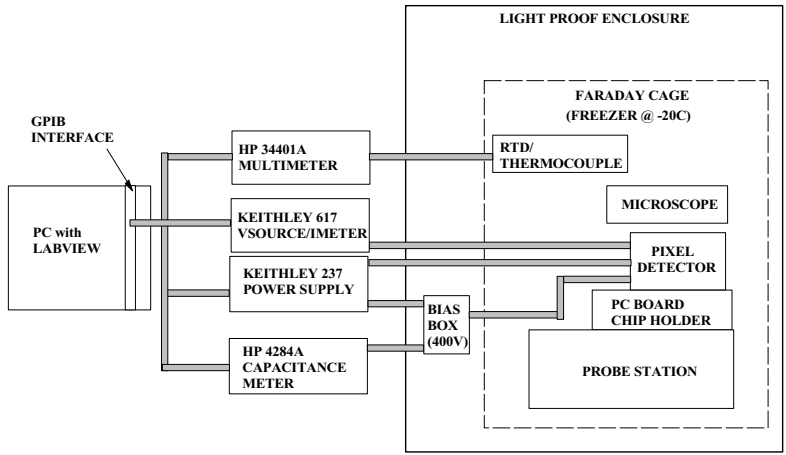

Figure 8: Block diagram of the probe station setup.

# 4 Results

## 4.1 LBNL Test Structure Studies

### 4.1.1 Inter-pixel Capacitance of the LBNL Test Structures

Measurements were made on unirradiated and irradiated ($4.8 \times 10^{13}$ cm$^{-2}$ 1 MeV neutron equivalent fluence) LBNL structures at frequencies ranging from 3 kHz to 1 MHz. Figures 9 and 10 show the capacitance versus voltage for the unirradiated $n$-type and $p$-type detectors, respectively, and Tables 3 and 4 summarize the capacitances at full depletion. No variation with frequency was seen in capacitance above full depletion. In the low voltage regime of the $n$-type detector's response, the 1 MHz data lie about 15% below the values at the other frequencies. This phenomenon was not investigated further as it does not affect the behavior of the detectors, since they are normally operated at higher voltages even if underdepleted.

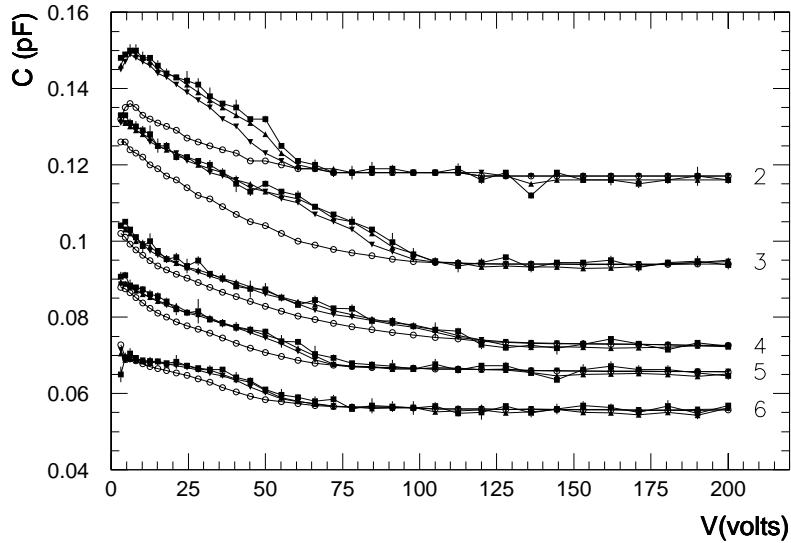

Figure 9: Inter-pixel capacitance of an unirradiated $n$-type LBNL detector. The multiple curves represent measurements made at frequencies 3, 10, and 100 kHz and 1 MHz. The families of curves labelled 2–6 show measurements of the arrays with the corresponding numbers (see Table 1 for their characteristics). The open circles are the 1 MHz data.

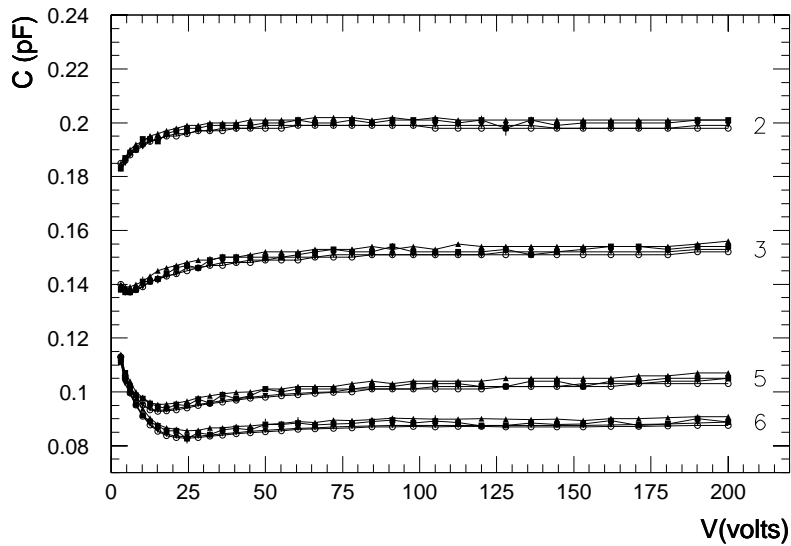

Figure 10: Inter-pixel capacitance of an unirradiated $p$-type LBNL detector. The multiple curves represent measurements made at frequencies 3, 10, and 100 kHz and 1 MHz. The families of curves labelled 2, 3, 5, and 6 show measurements of the arrays with the corresponding numbers (see Table 1 for their characteristics).

Figures 11 and 12 show the capacitance as a function of voltage for an irradiated $n$-type and $p$-type detector, respectively. The results are consistent with those reported for a similar measurement[4] of silicon strip sensors; in particular, after irradiation the dependence of the capacitance upon the bias voltage is significantly increased. Thermal runaway at high voltages in these irradiated structures prevented data from being taken above 200 V at room temperature.

As the capacitance appears still to be decreasing at the highest voltage measured, the endpoint of the data is likely to overestimate the minimum. The curves are well described by an exponential function plus a constant offset; it is possible, however, that the capacitance could flatten out more abruptly than this at higher voltages. To estimate the capacitance at a voltage corresponding to complete depletion of the volume between the implants, we average the endpoint of the data with the asymptotic value of the exponential plus constant function. The extrapolation error is taken to be half the difference between the two values. These values are listed in Tables 3 and 4.

The capacitance of the irradiated sensors was seen to depend more strongly on the signal frequency than was the case for the unirradiated detectors. This frequency dependence phenomenon has been described elsewhere[7, 8] for backplane capacitance and so was not explored in this study. As the 1 MHz data are most relevant to planned or likely future readout systems at LHC and Tevatron experiments, only those data

**Table 3** Measured inter-pixel capacitance of unirradiated and irradiated (4.8 × $10^{13}$ n/cm$^2$ fluence) $n$-type LBNL detectors. The measurements of the unirradiated detectors are shown in Figure 9. The measurements of the irradiated detectors are shown in Figure 11 and have been extrapolated beyond full depletion as is explained in the text. The average of the extrapolated value and the endpoint is listed in Column 6. The total error on each measurement is conservatively estimated as the quadrature sum of the extrapolation error (Column 7) and the combination of the statistical and systematic errors, which is 5 fF.

| Array No. | $p^+$ Implant Width ($\mu$m) | $C^{\text{unirrad}}$ (fF) | $C^{\text{irrad}}_{\text{endpoint}}$ (fF) | $C^{\text{irrad}}_{\text{extrap.}}$ (fF) | $C^{\text{irrad}}_{\text{average}}$ (fF) | Extrap. Error (fF) | Tot. Error on $C^{\text{irrad}}$ (fF) |
|---|---|---|---|---|---|---|---|
| 2 | 38 | 115 | 118 | 110 | 114 | 4 | 7 |
| 3 | 32 | 94  | 98  | 94  | 96  | 2 | 6 |
| 4 | 23 | 73  | 75  | 66  | 71  | 5 | 7 |
| 5 | 20 | 66  | 71  | 61  | 66  | 5 | 7 |
| 6 | 14 | 56  | 60  | 51  | 56  | 5 | 7 |

**Table 4** Measured inter-pixel capacitance of unirradiated and irradiated (4.8 × $10^{13}$ n/cm$^2$ fluence) $p$-type LBNL detectors. The measurements of the unirradiated detectors are shown in Figure 10. The measurements of the irradiated detectors are shown in Figure 12 and have been extrapolated beyond full depletion as is explained in the text. The average of the extrapolated value and the endpoint is listed in Column 6. The total error on each measurement is conservatively estimated as the quadratic sum of the extrapolation error (Column 7) and the combination of the statistical and systematic errors, which is 5 fF.

| Array No. | $n^+$ Implant Width ($\mu$m) | $C^{\text{unirrad}}$ (fF) | $C^{\text{irrad}}_{\text{endpoint}}$ (fF) | $C^{\text{irrad}}_{\text{extrap.}}$ (fF) | $C^{\text{irrad}}_{\text{average}}$ (fF) | Extrap. Error (fF) | Tot. Error on $C^{\text{irrad}}$ (fF) |
|---|---|---|---|---|---|---|---|
| 2 | 38 | 200 | 218 | 217 | 218 | 1  | 5  |
| 3 | 32 | 153 | 166 | 152 | 159 | 7  | 9  |
| 5 | 20 | 103 | 123 | 108 | 116 | 8  | 10 |
| 6 | 14 | 88  | 112 | 87  | 100 | 13 | 14 |

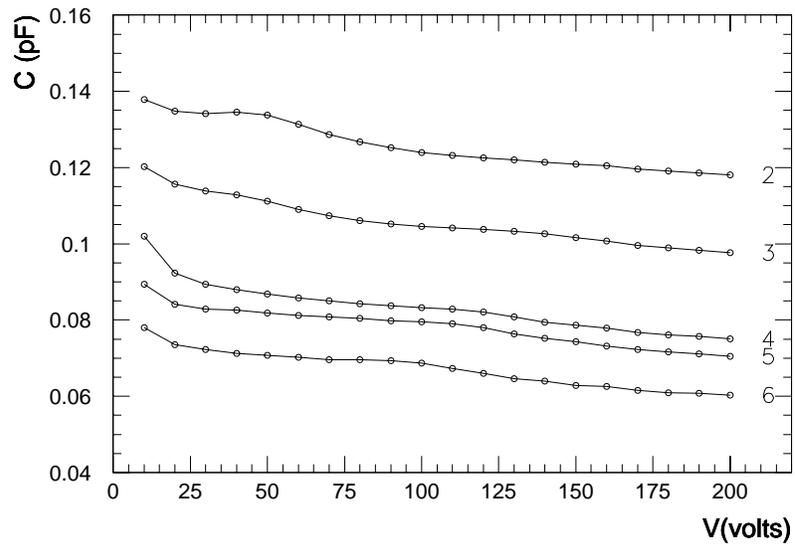

Figure 11: Inter-pixel capacitance of an irradiated ($4.8 \times 10^{-13}$ 1 MeV neutron equivalent) $n$-type LBNL detector. The families of curves labelled 2–6 show measurements of the arrays with the corresponding numbers (see Table 1 for their characteristics). Measurements were taken at 1 MHz.

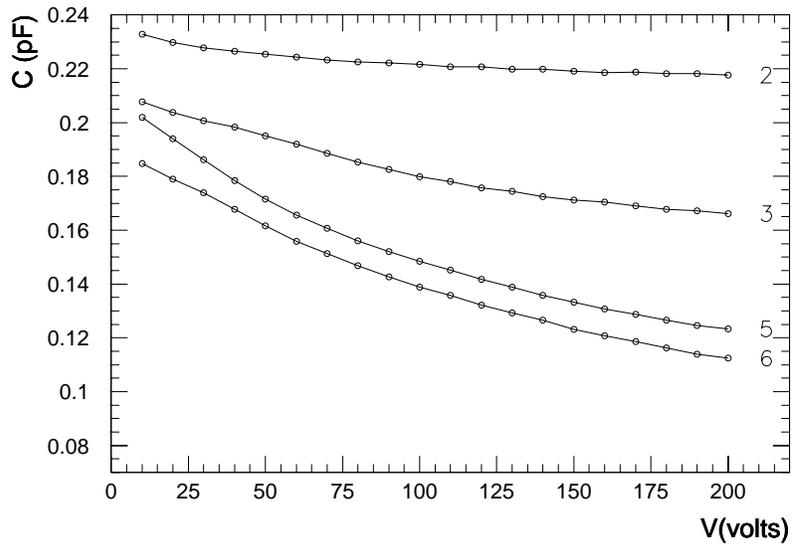

Figure 12: Inter-pixel capacitance of an irradiated ($4.8 \times 10^{-13}$ 1 MeV neutron equivalent) $p$-type LBNL detector. The families of curves labelled 2, 3, 5, and 6 show measurements of the arrays with the corresponding numbers (see Table 1 for their characteristics). Measurements were taken at 1 MHz.

are shown.

We have investigated whether the capacitance varies linearly with the width-to-pitch ratio ($W/\wp$) as has been shown [9] to be the case for silicon strip sensors. We take $50\mu$m to be the relevant pitch in this study as the capacitance between the short sides of adjacent pixels (Cd or Ce in Figure 16) is only about 10% of the capacitance between the long sides (Cp). Because the pitch is the same in all of the sensors that we studied, we are not able to investigate explicit dependence of capacitance upon pitch but instead test whether the relationship to the width is linear. Figure 13 shows that a linear fit describes the data adequately up to a width of $32\mu$m. This is consistent with the validity region ($W/\wp < 0.6$) given in Reference [9]. However, a linear relationship fails to describe the data for larger implant widths in $p$-type sensors. Figure 13 shows that an improved fit is achieved for the function

$$C_{\text{inter-pixel}} = A + BW/\wp + D/G, \tag{1}$$

where $W$, $\wp$, and $G$ are the width, pitch, and gap, respectively, as illustrated by Figure 1 and listed in Table 1. For the LBNL test structures, the p-stop width is proportional to the gap width, so $G$ is proportional to $50 - W$; hence it is possible to plot the result as a function of $W$. Lines 1–5 of Table 5 summarize the results of the fit.

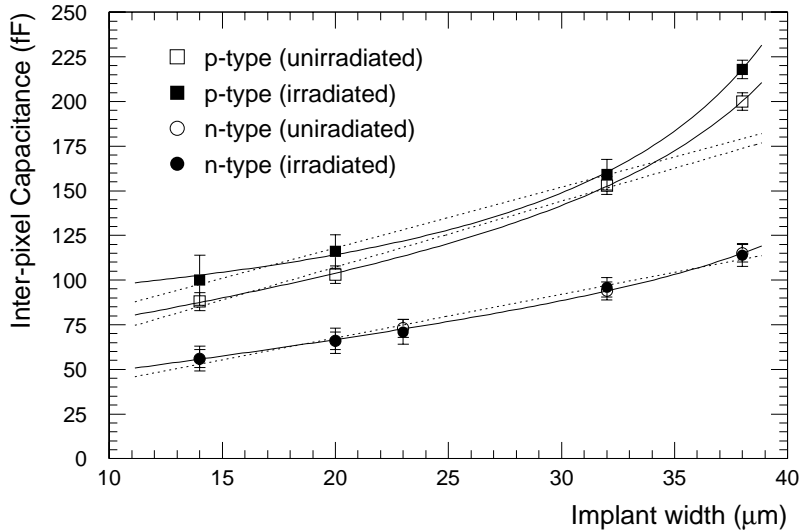

Figure 13: Inter-pixel capacitance versus width for irradiated ($4.8 \times 10^{13}$ n/cm$^2$ fluence) and unirradiated LBNL $n$- and $p$-type sensors of pitch $50\mu$m. Fits to a straight line and to the function given in Equation (1) are shown.

**Table 5** Summary of fit parameters found to describe the inter-pixel capacitance of unirradiated sensors. The fit function is $C_{\text{inter-pixel}} = A + BW/\wp + D/G$, where $G$ is the total unimplanted gap between pixels in microns, $W$ is the pixel implant width, $\wp$ is the pitch (50$\mu$m for all sensors), and $C_{\text{inter-pixel}}$ is the inter-pixel capacitance per unit length in pF/cm. (The length was chosen to be the pitch in the long dimension of the pixels—536$\mu$m for the LBNL test structures and 400$\mu$m for the Structure 6 devices.) Fits with and without the third term were examined. $N$ is the number of degrees of freedom per fit. "Str 6, Set 1" is a fit to all CiS devices and Seiko devices (see Section 4.2.1) utilizing the "Atoll" p-stop design. "Str 6, Set 2" is a fit to Seiko devices with "Common" and "Combined" p-stop designs.

|  | A (pF/cm) | B (pF/cm) | D ($\mu$m·pF/cm) | $\chi^2/N$ |
|---|---|---|---|---|
| LBNL $p$-type | $0.63 \pm 0.17$ | $3.44 \pm 0.36$ | – | 1.1[†] |
|  | $0.32 \pm 0.14$ | $4.26 \pm 0.25$ | – | 5.4 |
|  | $0.53 \pm 0.15$ | $1.34 \pm 0.92$ | $17.5 \pm 5.3$ | 0.058 |
| LBNL $n$-type | $0.35 \pm 0.13$ | $2.28 \pm 0.25$ | – | 0.48 |
|  | $0.43 \pm 0.15$ | $1.32 \pm 0.85$ | $8.5 \pm 7.2$ | 0.015 |
| Str 6, Set 1 | $0.42 \pm 0.09$ | $2.70 \pm 0.21$ | – | 1.1 |
|  | $0.40 \pm 0.09$ | $1.61 \pm 0.43$ | $8.2 \pm 2.8$ | 0.87 |
| Str 6, Set 2 | $0.23 \pm 0.14$ | $1.77 \pm 0.29$ | – | 0.35 |
|  | $0.16 \pm 0.16$ | $1.31 \pm 0.52$ | $4.3 \pm 4.0$ | 0.27 |

[†] Fit to first three points only (i.e., for $W \leq 32$ $\mu$m).

The function in Equation (1) may be interpreted as follows. The $1/G$ term describes the contribution to the capacitance from regions near the sensor surface, while the $W/\wp$ dependence describes the contribution from regions deeper in the bulk. The $W/\wp$ component may be expected if one approximates the two neighbors as a parallel plate capacitor (capacitance $C_1$ in Figure 1) of width $W$ and average distance between plates proportional to the pitch. In the region close to the surface, the capacitance is dominated by the capacitances between the implants (capacitances $C_2$ in Figure 1). Because the p-stops, if present, provide low resistance, the effective distance between the plates is determined by the unimplanted gap. In this region the effect of the implant width is relatively unimportant. This combination leads to the term proportional to $1/G$. The contribution from pixels other than the nearest neighbors is not expected to have significant dependence on $W$ [10] and so enters principally through the constant term. We note that the second and third term on the righthand side of Equation (1) are strongly coupled—that is, the division between regions near and far from the surface is not sharp.

### 4.1.2 Backplane Capacitance of the LBNL Test Structures

The setup for the backplane capacitance measurement is similar to that for the inter-pixel measurement with the LOW terminal connected to the center pixel. The HIGH terminal, however, is not connected to the surrounding and neighboring pixels but is connected to the backside of the sensor. Measurements were made on unirradiated LBNL $n$- and $p$-type pixels. The results are summarized in Tables 6 and 7. Figure 14 shows a typical measurement. We have chosen not to report extensively on the backplane capacitance of irradiated pixel sensors because the radiation response of the bulk from which it is derived has been examined by us previously[1] and found to be well described by existing models.

**Table 6** Measured and simulated (see Section 4.1.3) values of the backplane capacitance of an LBNL $n$-type detector. The total error on each measurement is conservatively estimated at 5fF. The uncertainty on the simulation results is $\pm 20\%$.

| Array Number | Measured C (fF) | Simulated $C_{IES2D}$ (fF) | Simulated $C_{IES3D}$ (fF) |
|---|---|---|---|
| 2 | 15 | 10.4 | 13.4 |
| 3 | 15 | 10.3 | 13.1 |
| 4 | 11 | 10.3 | 12.7 |
| 5 | 15 | 10.3 | 12.6 |
| 6 | 13 | 10.2 | 12.2 |

**Table 7** Measured values of the backplane capacitance of an LBNL $p$-type detector. The total error on each entry is conservatively estimated at 5fF.

| Array Number | Measured C (fF) |
|---|---|
| 2 | 18 |
| 3 | 16 |
| 4 | 11 |
| 6 | 21 |

### 4.1.3 Electrostatic Simulations of the LBNL Test Structures

Electrostatic simulations [11] of the inter-pixel and backplane capacitances of the LBNL $p^+$-on-$n$ devices were carried out to aid in the interpretation of the capacitance measurements. The agreement reported below between the simulation and the measurements confirms that the parasitic capacitances and the systematics associated with the measurement have been properly interpreted and addressed. The static calculations also help determine the frequency-independent capacitance values in the pixel structure. Simulations were conducted in two and three dimensions with different simulators to provide an indicator of the precision of the simulation and the magnitude of contributions of order higher than nearest neighbors.

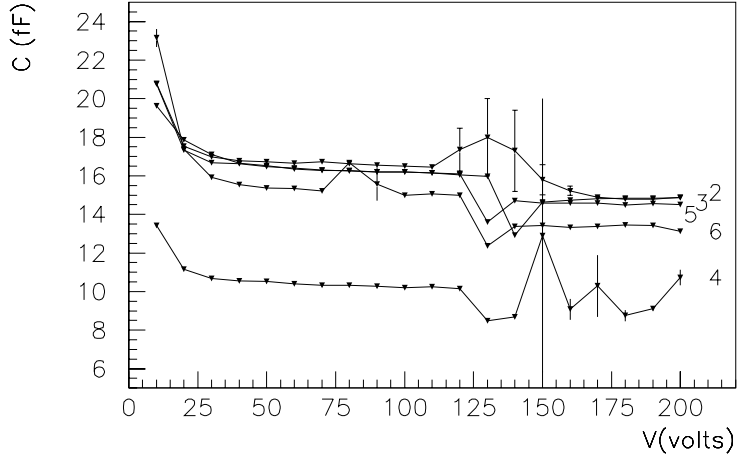

Figure 14: The backplane capacitance versus bias voltage for a typical LBNL $n$-type test structure. The families of curves labelled 2–6 show measurements of the arrays with the corresponding numbers (see Table 1 for their characteristics).

The two-dimensional electrostatic simulators, HSPICE [12] and IES Electro [13], take as input the geometrical dimensions of a vertical cross section of the pixel sensor and information on the dielectric medium (see Figure 15). They solve the electrostatic field equations of the defined design for capacitance. The results of these studies are shown in Tables 6 and 8. Table 9 shows the predicted contribution of the higher order parallel capacitance terms (Cp2 and Cp3 in Figure 15). The results of the two simulations bracket the measured values for the inter-pixel capacitance. A comparison of the two simulators leads us to assign an uncertainty of $\pm 35\%$ to the simulation result.

The calculation was repeated with the three-dimensional electrostatic field simulator IES Coulomb [14]. The three-dimensional model improves on the two-dimensional one by adding contributions to the capacitance associated with coupling to the implant's ends (Ce in Figure 16) and with diagonal couplings (Cd in the same figure). These contributions were then summed to predict the total inter-pixel capacitance associated with the nearest neighbors. These three-dimensional results are summarized in Tables 6 and 8 and agree with the measurements to an accuracy of better

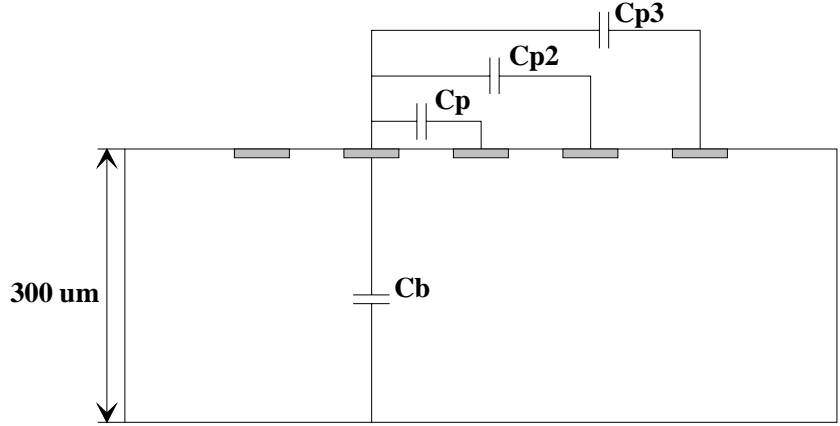

Figure 15: Geometry used to extract capacitance predictions for two-dimensional simulations of the pixel array structure.

Table 8 Simulated values of inter-pixel capacitance of unirradiated $n$-type LBNL detectors.

| Array Number | $p^+$ Implant Width ($\mu$m) | Simulated $C_{\text{HSPICE}}^{\text{unirrad}}$ (fF) | Simulated $C_{\text{IES2D}}^{\text{unirrad}}$ (fF) | Simulated $C_{\text{IES3D}}^{\text{unirrad}}$ (fF) |
|---|---|---|---|---|
| 2 | 38 | 130 | 109 | 124 |
| 3 | 32 | 115 | 91 | 111 |
| 4 | 23 | 95 | 78 | 93 |
| 5 | 20 | 89 | 72 | 87 |
| 6 | 14 | 75 | 66 | 76 |

than 30%.

**Table 9** Higher order contributions to the inter-pixel capacitance of an LBNL $n$-type sensor as simulated with the IES Coulomb two-dimensional electrostatic field solver. See Figure 15 for explanations of the column headings.

| Array Number | Gap Size ($\mu$m) | Cp (fF) | Cp2 (fF) | Cp3 (fF) |
|---|---|---|---|---|
| 2 | 12 | 46 | 5.30 | 3.30 |
| 3 | 18 | 37 | 5.25 | 3.25 |
| 4 | 27 | 31 | 5.20 | 3.15 |
| 5 | 30 | 28 | 5.15 | 3.10 |
| 6 | 36 | 25 | 5.00 | 3.05 |

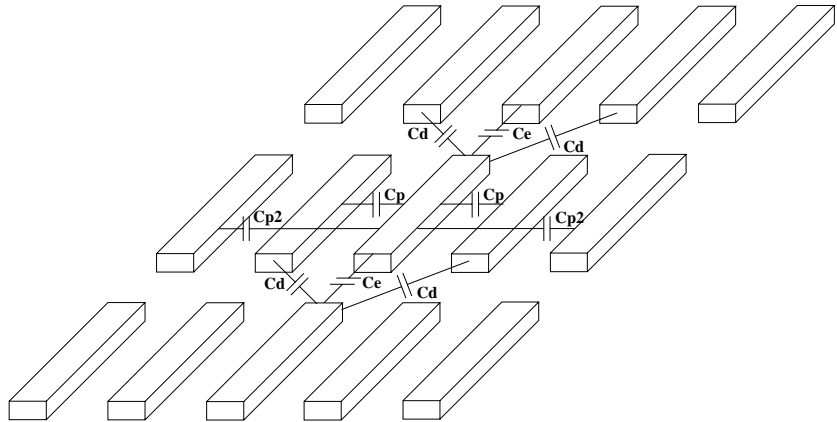

Figure 16: Geometry used for extracting capacitance predictions from a three-dimensional model of the test structure array.

### 4.1.4 Implications of LBNL Test Structure Simulations

The good agreement between the three-dimensional simulation results and the measured capacitance of the LBNL pixel devices gives us assurance that our test setup and measurement procedure can be used reliably for inter-pixel and backplane capacitance measurements in the range 10–100 fF. We conclude that this setup and procedure can be used for measuring Structure 6, devices described in Reference [5], and other pixel sensors of comparable geometry. The typical pixel designs represented in this study of the LBNL test structures indicate that the total pixel capacitance includes backplane and inter-pixel capacitances in a ratio of about 10–25% for both $p$-on-$n$ and $n$-on-$p$ devices.

## 4.2 Structure 6 Studies

### 4.2.1 Introduction

The first ATLAS pixel sensor prototypes were fabricated by two manufacturers, CiS (Erfurt, Germany) and Seiko (Chiba, Japan). The wafers bore a variety of test structures, including Structure 6, as well as two "tiles," which are sensors used in full-scale prototype ATLAS modules[15]. Array 1 within Structure 6 has the same implant dimensions and isolation technology (p-stops) as Tile 1.

Figure 17 shows the inter-pixel capacitance as a function of applied bias voltage, as measured on Array 1 from two Structure 6 devices on a CiS wafer and one on a Seiko wafer. Measurements were taken at five frequencies (1, 3, 10, and 100 kHz, and 1 MHz), and no frequency dependence was seen. (The 1 kHz measurements are consistent with the others but are somewhat noisier; they are omitted from Figure 17 for clarity.) The CiS measurements showed significant variation with applied bias, their curves of capacitance versus voltage ("$C_{inter-pixel}$-V") not flattening out until about 400 V. Measurements taken on Seiko devices showed little variation with applied bias voltage above full depletion. The principal difference between devices from the two vendors is the crystal orientation used ($\langle 100 \rangle$ for Seiko and $\langle 111 \rangle$ for CiS). As the crystal orientation affects the interface states, it is a possible cause of the difference in the characteristics. Other details of the processing may be involved as well.

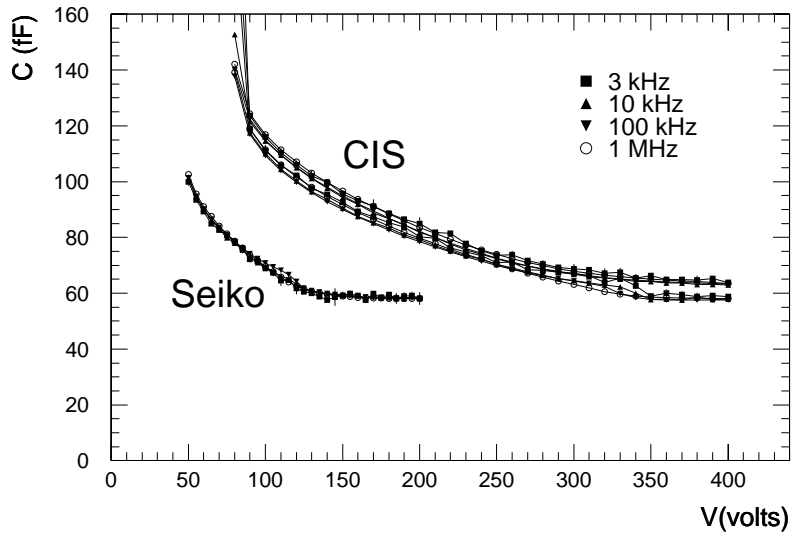

Figure 17: Inter-pixel capacitance as a function of applied bias voltage for Array 1 of Structure 6. Measurements from two devices from a CiS wafer and one representative one from a Seiko wafer are shown.

### 4.2.2 P-stop Design Dependence of Inter-pixel Capacitance for Unirradiated Sensors

Capacitance measurements were taken of pixels in each of the 11 arrays. Figure 18 shows data taken at 1 MHz at voltages for which the $C_{inter-pixel}$-V curves are flat (400 V for the CiS structures and 200 V for the Seiko ones). As was summarized in Table 2, Arrays 1–5 use the atoll p-stop design, Array 6 uses the combined p-stop, and Arrays 7–11 use the common p-stop. Array 2 has the same implant shape as Array 1, but, unlike Array 1, its metal is wider than its $n^+$ implant. (Having its metal wider than its implant alters the potential near the surface and, as is shown below, leads to increased inter-pixel capacitance.)

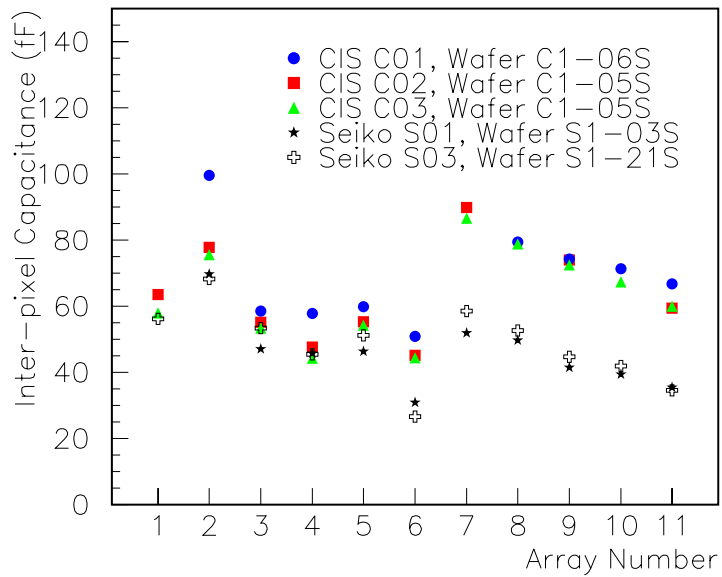

Figure 18: Inter-pixel capacitance of the 11 arrays on Structure 6 devices. See Table 2 for details on the design of each array.

We find that the capacitance of arrays with atoll p-stop design is relatively insensitive to the specific values of the implant dimensions for the design variations studied here. (The constraint associated with fitting an atoll p-stop into a 50$\mu$m pitch limited the range of $n^+$ implant widths that could be studied.) The capacitance of the common p-stop design shows the greatest variation with implant shape. As is shown in the figure, the capacitance of the five common p-stop designs decreases monotonically with array number. Arrays 7, 8, and 9 have the same gap width but increasing p-stop width. Arrays 9, 10, and 11 have the same p-stop width but increasing gap width. The contributions of the gap width and the p-stop width to the inter-pixel capacitance appear to be comparable. The combined p-stop design has the smallest $n^+$ implant width (and consequently the largest distance between $n^+$ implants) and was observed to have the smallest capacitance of the designs tested.

### 4.2.3 P-stop Design Dependence of Inter-pixel Capacitance for Irradiated Sensors

Several of the Structure 6 devices were irradiated beyond inversion by a fluence equivalent to $8 \times 10^{14}$ cm$^{-2}$ 1 MeV neutrons. Figure 19 shows the 1 MHz inter-pixel capacitance of two representative CiS devices as a function of voltage, both before and shortly after irradiation (i.e., after beneficial annealing but before significant onset of reverse annealing). Post-annealing data are not available for these devices. We find that the inter-pixel capacitance of the devices is initially increased by irradiation, even for devices beyond inversion.

Figure 20 shows the capacitance before and shortly after irradiation of all the arrays in a Seiko and a CiS sensor. While the values after irradiation are systematically higher than before, the size of the difference is consistent with the resolution on the measurement.

### 4.2.4 Inter-pixel Capacitance Versus Width-to-pitch Ratio

As can be seen in Figure 21, a linear relationship between the inter-pixel capacitance and the width describes the data well, although some minor inconsistencies are observed. Of note is that Array 10 has slightly wider $n^+$ implants than Array 9 yet consistently shows a lower capacitance. Array 10 has wider gaps, so this result suggests that the size of the unimplanted gap plays a roll in the capacitance. This effect is described if one instead fits the data to Equation (1). The results of the fitting procedure are summarized in lines 6–9 of Table 5.

While the data for all the p-stop designs in CiS structures are well described by a single parameterization (for either a straight line or Equation (1)), the Seiko structures show two groupings. The common and combined p-stop designs form one group whose capacitances are lower than those of the CiS structures. The atoll p-stop structures form a separate group with capacitances more consistent with those of the CiS structures. The origin of this effect is not yet understood.

### 4.2.5 Temperature Dependence of Capacitance

The silicon pixel detectors in ATLAS will operate at a temperature of −6°C. Other future pixel systems are likely to utilize low temperatures as well. Because the high leakage currents that are present after irradiation limit the voltage that can be applied, low temperature operation allows measurements to be made over a larger range of operating voltages. To quantify the temperature dependence of the inter-pixel capacitance, measurements were made at various temperatures from room temperature to below freezing.

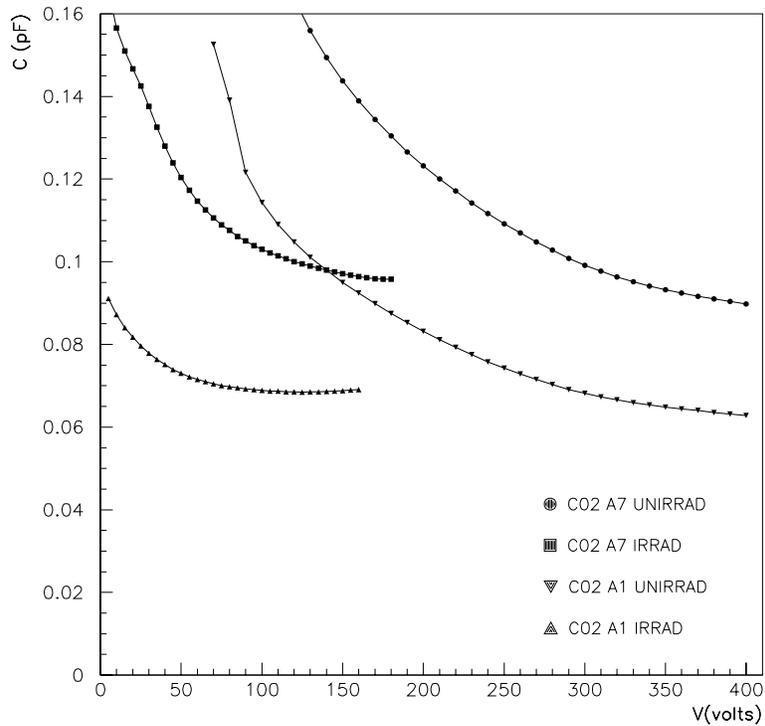

Figure 19: Inter-pixel capacitance of several Structure 6, Array 1 devices from the same wafer, before and after irradiation to a fluence equivalent to $8 \times 10^{14}$ cm$^{-2}$ 1 MeV neutrons. The combined systematic and statistical uncertainty on any particular measurement is 5 fF. Only the 1 MHz data are shown.

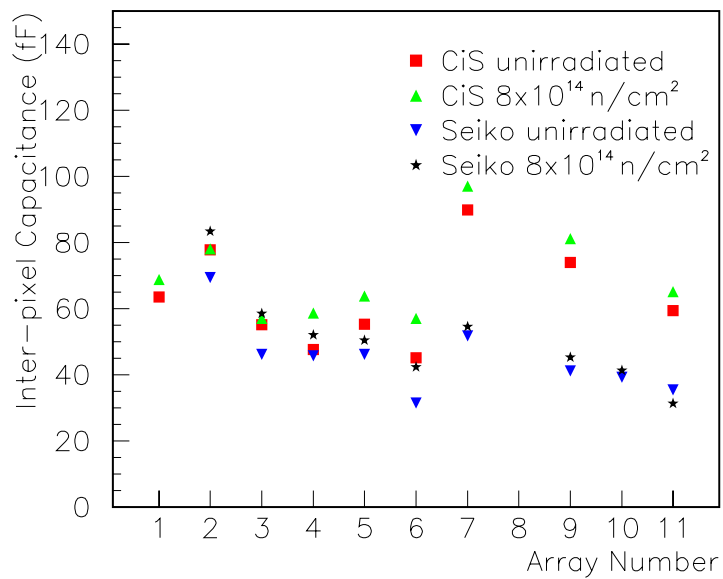

Figure 20: Inter-pixel capacitance of the 11 arrays in one Seiko and one CiS Structure 6 device before and shortly after irradiation. The combined statistical and systematic error on each point is 5 fF.

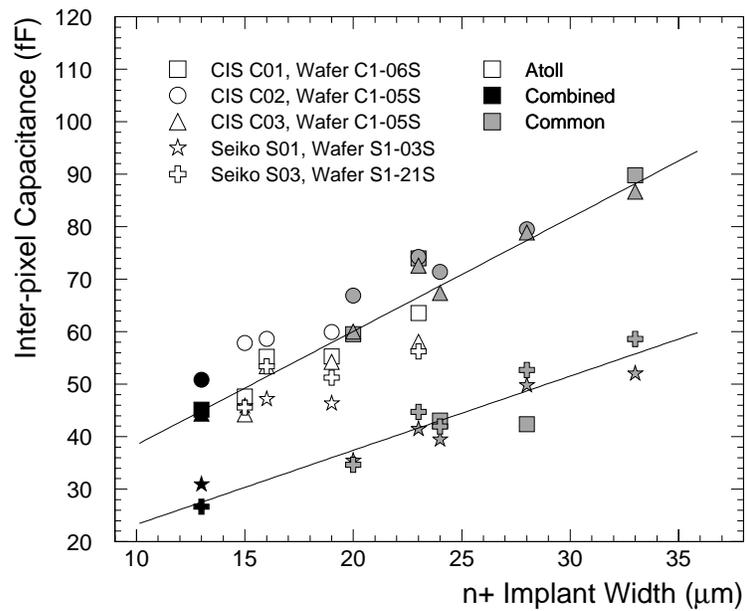

Figure 21: Inter-pixel capacitance versus implant width for unirradiated Structure 6 sensors of pitch 50$\mu$m. The combined statistical and systematic error on each point is 5 fF.

The variation in temperature was achieved by placing the probe station in a cooling chamber which was initially off and at room temperature. The chamber was turned on and measurements were made as the sensor cooled down. The rate of cooling was slow enough to allow the sensor to equilibrate between measurements. Reliable measurements were possible down to a temperature of about $-10°C$. Figure 22 shows the capacitance measured at 1 MHz frequency versus temperature for several Structure 6 devices and LBNL test structures. The results show that in the range $+20°C$ to $-10°C$, capacitance measured at this frequency is temperature-independent. The temperature range and frequency value used in this measurement complement and extend those reported in two other studies [8, 16].

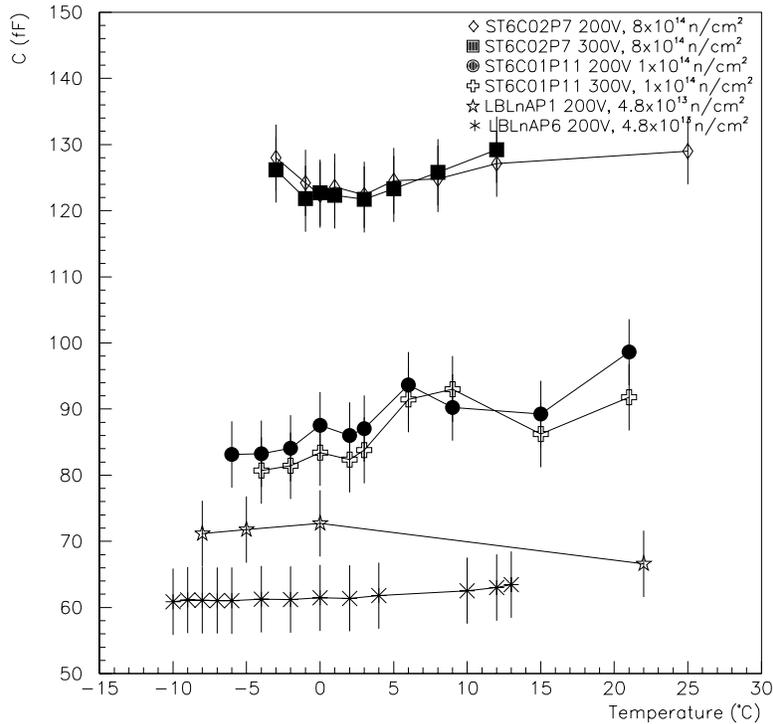

Figure 22: Inter-pixel capacitance of Structure 6 devices and LBNL test structures, measured as a function of temperature.

# Acknowledgments

This study would not have been possible without the assistance of Stephen Holland, who designed the LBNL test structures and offered valuable advice about the interpretation of the data. We also gratefully acknowledge the assistance of Carl Allen, who fabricated custom fixturing and equipment. Murdock Gilchriese provided essential support for the irradiations, and Alexander Brandl, Gavin Mendel-Gleason, and Trek Palmer provided several drawings and figures. Anthony Riggins assisted with the electrostatic simulation. This work was supported in part by the U.S. Department of Energy. The participation by the undergraduate students was made possible by a NASA PURSUE Partnership Award for the Integration of Research.